\documentclass{article}

\usepackage{mathrsfs}
\usepackage{amssymb}
\usepackage{epsfig}

\newcounter{figuur}

\newtheorem{theorem}{Theorem}
\newtheorem{lemma}[theorem]{Lemma}

\newtheorem{corollary}[theorem]{Corollary}
\newcommand{\s}{\sigma}
\newcommand{\OO}{\Omega}
\newcommand{\half}{{\textstyle \frac{1}{2}}}
\newcommand{\fourth}{{\textstyle \frac{1}{4}}}
\newcommand{\CH}{\mathcal{H}}
\newcommand{\CF}{\mathcal{F}}

\newcommand{\CS}{\mathcal{S}}
\newcommand{\CB}{\mathcal{B}}
\newcommand{\CA}{\mathcal{A}}
\newcommand{\CC}{\mathcal{C}}
\newcommand{\CZ}{\mathcal{Z}}
\newcommand{\spec}{\mathrm{\textbf{Spec}}}
\newcommand{\ten}{\otimes}
\newcommand{\ket}[1]{|#1\rangle}
\newcommand{\bra}[1]{\langle#1|}
\newcommand{\ketbra}[1]{\ket{#1}\bra{#1}}
\newcommand{\inp}[2]{\langle #1,#2\rangle}
\newcommand{\one}{\mathbf{1}}
\newcommand{\tr}{\mathbf{tr}}

\newcommand{\up}{\ket{\! \uparrow \,}\bra{\, \uparrow \!}}
\newcommand{\down}{\ket{\! \downarrow \,}\bra{\, \downarrow \!}}
\newcommand{\fig}{\refstepcounter{figuur}\textbf{Fig.\,\arabic{figuur}: }}

\newcommand{\qed}{\penalty10000\hfill$\Box$\par\goodbreak\medskip}

\newcommand{\proof}{\medskip\noindent{\it Proof:} }
\newcommand{\definition}{${}$\\[0.3 cm]\noindent{\bf Definition.} }

\begin{document}

\title{Unifying Decoherence and the Heisenberg Principle}
\author{Bas Janssens}
\maketitle
\abstract{We exhibit three inequalities
involving quantum measurement, all of which are 
sharp and state independent.
The first inequality bounds the performance of joint measurement.
The second 
quantifies the trade-off between
the measurement quality and
the disturbance caused on the measured system.  
Finally, the third inequality provides a sharp 
lower bound on the amount of decoherence
in terms of the measurement quality.  
This gives a unified description of 
both the Heisenberg uncertainty principle and 
the collapse of the wave function. 
}
\section{Introduction}
In quantum mechanics, observables are modelled by self-adjoint 
operators $A$, and states by normalized trace-class operators $\rho$. 
A state $\rho$ induces a probability 
measure on an observable $A$. 
It is the objective of a \emph{quantum measurement}
to portray this probability measure as 
faithfully as possible.

According to the uncertainty relation  
$
\sigma_{X} \sigma_{Y}
\geq \half |\tr (\rho [X,Y])|\,,
$
(see \cite{He, Ke, Ro}),
there is an inherent variance 
in the quantum state.
Furthermore, quantum theory puts severe 
re\-stric\-tions on the performance of measurement.
These restrictions, which come on top of the measurement restrictions
implied by the above uncertainty relation, fall into three distinct classes.
\begin{itemize}
\item[I]The impossibility of perfect \emph{joint measurement}.
It is not possible to perform a simultaneous measurement
of two noncommuting observables in such a way that 
both measurements have perfect quality. 
\item[II]The \emph{Heisenberg principle}, (see \cite{He}).
This states that 
quantum information cannot be extracted from a system without disturbing 
that system. 
\item[III]The \emph{collapse of the wave function}.
When information is extracted from a quantum system, 
a so-called decoherence is experimentally known to 
occur on this system. 
\end{itemize}
We will see that this collapse of the wave function is
a mathematical consequence of information extraction.
In the process, II and III will be clearly exhibited as 
two sides of the same coin.

The subject of uncertainty relations in quantum measurement is already 
endowed with an extensive literature.
For example, 
the Heisenberg principle and the impossibility of joint measurement
are quantitatively illustrated in
\cite{AK,Oz, Is, Ha}.

However, 
the inequalities in these papers depend on the state $\rho$,
which somewhat limits their practical use.
Indeed, the bound on the measurement quality 
can only be calculated if the state $\rho$
is known, in which case there is no need for a 
measurement in the first place.

Our state-independent figures of merit  
(sections \ref{trans} and \ref{mav}) 
will lead us quite naturally to 
state-independent bounds on the performance of measurement.
In order to illustrate their practical use,
we will give some applications. 
We investigate the beamsplitter, resonance fluorescence and 
nondestructive qubit measurement.

In section \ref{JM}, we will prove a sharp, state independent bound on 
the performance of jointly unbiased measurement. This generalizes the
impossibility of perfect joint measurement.

In section \ref{Heis}, we will prove a sharp, state independent bound on the 
performance of a measurement in terms of the 
maximal disturbance that it causes.
This generalizes the Heisenberg principle.
  
In contrast with the Heisenberg principle 
and its abundance of inequalities,
the phenomenon of decoherence has mainly been 
investigated in specific examples (see e.g. \cite{Hp,Zu,JZ}).
Although there are some bounds on the 
remaining coherence in terms of the measurement quality
(see \cite{JM,Se}),
a sharp, information-theoretic inequality does not yet appear to exist.
 
We will provide such an inequality in 
section \ref{collapse}, where
we will prove a sharp upper bound 
on the amount of coherence 
which can survive information transfer.
Not only does this generalize the collapse of the wave function, 
it also shows that no information can be extracted
if all coherence is left perfectly intact.
It is therefore a \emph{unified} description of both 
the Heisenberg principle and the collapse of the wave function.

\section{Information Transfer}\label{trans}

In quantum mechanics, a system is described by a von Neumann algebra $\CA$
of bounded operators on a Hilbert space $\CH$.  
(Usually the algebra $B(\CH)$ of all bounded operators.)
Its state space is formed
by the normalized density matrices $\CS(\CA) = \{ \rho \in
\CA  \, ; \, \rho \geq 0, \, \tr ( \rho ) = 1\}$. 
With the system in state $\rho \in \CS (\CA)$,
observation of a (Hermitean) observable $A \in \CA$ 
is postulated to yield the average value $\tr(\rho A)$.  
\definition Let $\CA$ and $\CB$ be von Neumann algebras.
A map $T : \CB \rightarrow \CA$ is called \emph{Completely Positive}
(or CP for short) if it is linear, normalized 
(i.e. $T(\one) = \one$), positive (i.e. $T(X^{\dag}X) \geq 0$ for all $X \in \CB$) and
if moreover the extension 
${\it Id}_{n} \ten T : M_n \ten \CB \rightarrow M_n \ten \CA$
is positive for all $n \in {\mathbb{N}}$, where $M_n$ is the algebra of
complex $n \times n$-matrices. In this article, we will require CP-maps 
to be weakly continuous unless specified otherwise.
\\[0.3 cm]
\noindent Its dual $T^{*} : \CS(\CA) \rightarrow \CS(\CB)$, defined by the
requirement $\tr(T^*(\rho) X) = \tr(\rho T(X)) \, \forall \, X \in \CB$,
has a direct
physical interpretation as an operation between quantum systems.
First of all, due to positivity and normalization of $T$, each state $\rho
\in \CS(\CA)$ is again mapped to a state $T^* (\rho) \in \CS (\CB)$. 
Secondly, linearity implies that $T^*$
satisfies $p T^* (\rho_1) + (1-p) T^*(\rho_2) =  T^* (p \rho_1
+ (1-p) \rho_2)$ for all $p \in [0,1]$, $\rho_1 , \rho_2 \in \CS(\CA)$. 
This expresses the 
\emph{stochastic equivalence principle}: a system which is in state $\rho_1$
with probability $p$ and in state $\rho_2$ with probability $(1-p)$ cannot
be distinguished from a system in state  $p \rho_1 + (1-p) \rho_2$.
Finally, it is possible to extend the systems $\CA$ and $\CB$
under consideration with another system $M_n$, on which the operation
acts trivially.  
Due to complete positivity, states in $\CS (M_n \ten \CA)$ are once again 
mapped to states in
$\CS(M_n \ten \CB)$.
Incidentally, any CP-map $T$ automatically satisfies $T(X^{\dag}) =
T(X)^{\dag}$ and $\|T(X)\| \leq \|X\|$ for all $X \in \CB$.

\subsection{General, Unbiased and Perfect Information Transfer}

Suppose that we are interested in the distribution of 
the observable $A \in \CA$, 
with the system $\CA$ in some unknown state $\rho$. We perform the
operation $T^* : \CS(\CA) \rightarrow \CS(\CB)$, and then observe 
the `pointer' $B$ in $\CB$ in order to obtain information on $A$. 
One may (see \cite{Ha}) 
take the position
that \emph{any} CP-map 
$T : \CB \rightarrow \CA$
is an information transfer 
from \emph{any} observable $A \in \CA$ to \emph{any}
pointer $B \in \CB$.
The following is a figure of 
demerit for the quality of such an
information transfer. 
\definition 
Let $T : \CB \rightarrow \CA$ be a CP-map. Its
\emph{measurement infidelity} $\delta$ in transferring information 
from $A$ to the pointer $B$ is defined as
$ \delta := \sup_{S} \| \one_{S}(A) - T(\one_{S}(B)) \|$,
where $S$ runs over the Borel subsets of $\mathbb{R}$. 
\\[0.3 cm] \noindent
It
measures how accurately 
probability distributions on 
the measured observable $A$ are copied to the pointer $B$. 

The initial state $\rho$ defines a probability distribution 
$\mathbb{P}_{i}$ on the spectrum of $A$
by $\mathbb{P}_{i}(S) := \tr(\rho \one_{S}(A))$,
where $\one_{S}(A)$ denotes the spectral projection of $A$
associated to the set $S$. 
Similarly, the final state $T^*(\rho)$ defines a probability distribution 
$\mathbb{P}_{f}$ on the spectrum of $B$.
$\delta$ is now the maximum
distance between $\mathbb{P}_{i}$
and $\mathbb{P}_{f}$, where the maximum is taken over all initial states
$\rho$.
That is, $\delta = 
\sup_{\rho} D(\mathbb{P}_{i} , \mathbb{P}_{f} )$.
 
The \emph{trace distance} (a.k.a. variational distance or Kolmogorov distance) 
is defined as
$D( \mathbb{P}_{f} , \mathbb{P}_{i} ) $
$:= \sup_{S} \{| \mathbb{P}_{i}(S) 
- \mathbb{P}_{f}(S) |\}$,
the difference between the probability that 
the event $S$
occurs in the distribution $\mathbb{P}_{i}$ and the 
probability that it occurs in the distribution 
$\mathbb{P}_{f}$, for the worst case 
Borel set $S$. 
Writing out this definition, we see that indeed
$
\sup_{\rho} D(\mathbb{P}_{i} , \mathbb{P}_{f} ) =
\sup_{\rho, S} |\tr(\rho \one_{S}(A)) - \tr(\rho T(\one_{S}(B)))| =
\sup_{S} \| \one_{S}(A) - T(\one_{S}(B)) \| = \delta $.
The measurement infidelity $\delta$ is precisely 
the worst case difference between
input and output probabilities.

In this article, we will devote considerable attention to the class of 
unbiased information transfers.
\definition A CP-map $T : \CB \rightarrow \CA$ is called an \emph{unbiased
information transfer} from the Hermitean observable $A \in \CA$ to 
a Hermitean $B \in \CB$ if $T(B) = A$.
\\[0.3 cm] \noindent
Recall that we are interested in the distribution of $A$, 
with the system $\CA$ in some unknown state $\rho$. We perform the
operation $T^* : \CS(\CA) \rightarrow \CS(\CB)$, and then observe 
the `pointer' $B$ in $\CB$.
Since $\tr(T^*(\rho) B ) = \tr(\rho T(B)) $ by definition of the dual, 
and $\tr(\rho T(B)) = \tr(\rho A)$ by definition of unbiased information transfer, 
the expectation value of $B$ in the final state $T^*(\rho)$ 
is the same as that of $A$ in the initial state $\rho$.  
We conclude that the expectation of $A$ was transferred to $B$. 
\definition An information transfer $T : \CB \rightarrow \CA$ 
from $A \in \CA$ to $B \in \CB$ is called \emph{perfect} if 
$T(B) = A$ and if
the restriction of $T$ to $B''$, the von Neumann algebra generated by $B$, 
is a  ${}^*$-homomorphism $B'' \rightarrow A''$.
\\[0.3 cm] \noindent
The entire probability distribution of $A$
is then transferred to $B$, rather than merely its average value. 
Indeed, for all moments
$\tr(\rho A^n)$, we have $\tr(T^*(\rho) B^n) = \tr(\rho T(B^n))
= \tr(\rho T(B)^n) = \tr(\rho A^n)$.  Everything there is to know about 
$A$ in the initial state $\rho$ can be obtained by observing the `pointer' $B$
in the final state $T^*(\rho)$. Note that the transfer is perfect
if and only if $\delta = 0$. 

Schematically, we have
\{General information transfers\} $\supset$
\{Unbiased information transfers\} $\supset$
\{Perfect information transfers\}.

\subsection{Example: von Neumann Qubit Measurement} \label{Neuqubit}
 
Let $\OO := \{ +1 , -1 \}$. 
Denote by $\CC(\OO)$ the (commutative) algebra
of $\mathbb{C}$-valued random variables on $\OO$. 
A state on $\CC(\OO)$ is precisely a probability distribution $\mathbb{P}$ on $\OO$,
and $\tr(\mathbb{P} f)$ should be read as  
$\mathbb{E}(f)$. 
Define the probability distributions $\mathbb{P}_{\pm}$ 
to assign probability 1 to $\pm 1$.
 
The Von Neumann-measurement $T : M_2 \ten \CC(\OO) \rightarrow M_2$
is defined as $T(X \ten f) := f(+1) P_+ X P_+ $ $+ f(-1) P_- X P_-$, with 
$P_+ = \up$ and $P_- = \down$.
Then $T^* : \CS(M_2) \rightarrow \CS(M_2) \ten \CS(\CC(\OO))$
is given by 
$T^* (\rho) = \tr(\rho P_+) \up \ten \mathbb{P}_+ + 
\tr(\rho P_-) \down \ten \mathbb{P}_-$.

In words: with probability $\tr(\rho P_+)$, the output $+1$
occurs and the qubit is left in \mbox{state $\ket{\!\uparrow\,}$}.  
With probability $\tr(\rho P_-)$, the output $-1$
occurs, leaving the qubit in state $\ket{\!\downarrow\,}$. 
The Von Neumann-measurement $T$ is a perfect 
(and thus unbiased) information 
transfer from $\s_z \in M_2$ to 
$\one \ten (\delta_{+1} - \delta_{-1}) \in M_2 \ten \CC(\OO)$. 

Quantum measurements are often (e.g. \cite{Ho,Ha}) modelled by 
Positive Operator Valued Measures or POVM's.
From our CP-map $T$, we may distill the POVM 
$\mu : \OO \rightarrow M_2$ by $\mu(\omega) := T(\one \ten
\delta_{\omega})$, i.e. $\mu(+1) = P_+$ and $\mu(-1) = P_-$.
This procedure is fully general: \emph{any} CP-map gives rise
to a POVM on a suitable $\Sigma$-algebra.  

A CP-map can thus be seen as an extension of a POVM that
keeps track of the system output as well as the measurement output.
Since we will be interested in disturbance of the system,
it is imperative that we consider the full CP-map rather than 
merely its POVM.

\section{Maximal Added Variance}\label{mav}
For unbiased information transfer, there exists a figure of demerit more
attractive than $\delta$.
Consider the variance $\mathbf{Var}(B,T^*(\rho))$ of the output.
(The variance is defined as 
$\mathbf{Var}(X,\rho) := \tr(\rho X^\dag X) - \tr(\rho X)^*
\tr(\rho X)$.)
The output variance can be split in two parts.
One part $\mathbf{Var}(A,\rho)$ is the variance of the input, which
is intrinsic to the quantum state $\rho$.
The other part $\mathbf{Var}(B,T^*(\rho))$ $-$
$\mathbf{Var}(A,\rho) \geq 0$ is \emph{added}
by the measurement procedure.
This second part determines how well the measurement performs.

The maximal added variance (where the maximum is taken over the
input states $\rho$) will be our figure of demerit.
For example, perfect information transfer from $A$ to $B$ 
satisfies
$\mathbf{Var}(B,T^*(\rho)) = \mathbf{Var}(A,\rho)$,
so that the maximal added variance is 0. 
There is uncertainty in the measurement outcome,
but all uncertainty `comes from' the quantum state, 
and none is added by the measurement procedure.
\definition The \emph{maximal added variance} of an 
unbiased information transfer $T$
is defined as
$$
\Sigma^2 := \sup_{\rho \in \CS(\CA)} \mathbf{Var}(B,T^*(\rho)) -
\mathbf{Var}(A,\rho)\,.
$$
It is straightforward to verify 
$\Sigma^2 = \| T(B^\dag B) - T(B)^\dag T(B) \|$. 
This inspires the following definition.
\definition Let $T : \CB \rightarrow \CA$ be a CP-map. 
We define the operator-valued 
sesquilinear form $( \,\cdot \,,\,\cdot \,) \, : \,  \CB \times \CB \rightarrow \CA$ by
$$(X,Y) := T(X^\dag Y) - T(X)^\dag T(Y)\,.$$ 
\noindent It satisfies $(X,Y)^\dag = (Y,X)$ and
is positive semi-definite: $(B,B) \geq 0$ for all $B \in \CB$.
This `length' has the physical interpretation $\|(B,B)\| = \Sigma^2$, 
and there is even a Cauchy-Schwarz inequality:
 \begin{lemma}[Cauchy-Schwarz] \label{causch}
 Let $T: \CB \rightarrow \CA$ be a CP-map, and
 $(X,Y) := T(X^\dag Y) - T(X)^\dag T(Y)$.  
 Then for all $X,Y \in \CB$:
 $$
 (X,Y)(Y,X) \leq \|(Y,Y)\| (X,X).
 $$
 \end{lemma}
 \proof By the Stinespring-theorem (see \cite{Ta}), we may assume 
 without loss of generality that
 $T$ is of the form $T(X) = V^\dag X V $ for some contraction $V$.
 Writing this out, we obtain 
 $(X,Y) = V^\dag X^\dag (\one - VV^\dag) YV$.
 Defining $g(X) := \sqrt{\one - VV^\dag} X V$, we write
 $(X,Y) = g(X)^\dag g(Y)$.
 Thus $(X,Y)(Y,X) = g(X)^\dag g(Y) g(Y)^\dag g(X) \leq 
 \|g(Y)\|^2 g(X)^\dag g(X) = \|(Y,Y)\| (X,X)$. \qed
\noindent 
If an information transfer is perfect, then of course 
$\Sigma^2 = \|(B,B)\| = 0$.
(No variance is added.) We will now show that the converse also holds: 
if $\Sigma^2 = 0$, then $T$ is a ${}^*$-homomorphism on
$B''$. (Compare this with the fact that probability distributions of 0
variance are concentrated in a single point.) 
\begin{theorem}\label{struc}
Let $T: \CB \rightarrow \CA$ be a CP-map, let $B \in \CB$ be Hermitean.
Then among
\begin{itemize}
\item[1] $(B,B) = 0$.
\item[2] The restriction of $T$ to $B''$, the von Neumann algebra generated
by $B$, is a ${}^*$-homomorphism $B''
\rightarrow T(B)''$. 
\item[3] $(f(B), f(B)) = 0$ for all measurable functions $f$ on the
spectrum of $B$.
\item[4] T maps the relative commutant $B' = \{X \in \CA ; [X,B] = 0\}$ 
into $T(B)'$.
\end{itemize}
the following relations hold: 
$(1) \Leftrightarrow (2) \Leftrightarrow (3) \Rightarrow(4)$. 
\end{theorem}
\proof  First $(1) \Rightarrow (2)$. 
By Cauchy-Schwarz (lemma \ref{causch}), we have
$T(B^n) - T(B)T(B^{n-1}) $ $\leq$ $\|(B,B)\| (B^{n-1}, B^{n-1}) $ $=$ $0$.
By induction, $T(B^n) = T(B)^n$, and by linearity $T(f(B)) = f(T(B))$
for all polynomials $f$.  Thus $T$ is a ${}^*$-homomorphism from the algebra of polynomials 
on the spectrum of $B$ to that on $T(B)$. Since $T$ is weakly continuous, this
statement extends to the algebras of measurable functions on the spectra of
$B$ and $T(B)$, isomorphic to $B''$ and $T(B)''$ respectively. 
For $(2)\Rightarrow(3)$, note that $T(f(B)^2) = T(f(B))^2$. 
For $(3)\Rightarrow(1)$, take \mbox{$f(x)=x$}. 
Finally we prove $(1)\Rightarrow(4)$: suppose that $[A,B] = 0$. Then 
$[T(B), T(A)] = T([A,B]) - [T(A), T(B)] = (A^\dag,B) - (B^\dag , A)$. 
($B$ is Hermitean.)  By Cauchy-Schwarz, the last term equals 
zero if $(B,B)$ does.
\qed
\noindent We see that the maximal added variance $\Sigma^2$ equals 0 if
and only if $T$ is a perfect information transfer.
We shall take $\Sigma$ to parametrize the imperfection of unbiased information
transfer.

\section{Joint Measurement}\label{JM}
In a jointly unbiased measurement, information on two observables $A$ and $\tilde{A}$ 
is transferred to two \emph{commuting} pointers $B$ and $\tilde{B}$.  
If $A$ and $\tilde{A}$ do not commute, then it is not possible for both
information transfers to be perfect. (See \cite{Ne}, \cite{We}.)
Indeed, the degree of imperfection is determined by the amount of
noncommutativity: 
\begin{theorem}\label{TJM}
Let $T: \CB \rightarrow \CA$ be a CP-map, let $B$,$\tilde{B}$ be commuting
Hermitean observables in $\CB$, and define $A := T(B)$, $\tilde{A} :=
T(\tilde{B})$, $\Sigma_{B}^2 := \|(B,B)\|$ and 
$\Sigma_{\tilde{B}}^2 := \|(\tilde{B},\tilde{B})\|$. Then 
\begin{equation}\label{jointje}
\Sigma_B \Sigma_{\tilde{B}} \geq \half \|[A,\tilde{A}]\| \,.
\end{equation}
\end{theorem}
\proof Since $[B, \tilde{B}] = 0$, we have 
$[\tilde{A}, A] = T([B,\tilde{B}]) - [T(B), T(\tilde{B})] = 
(B,\tilde{B}) - (\tilde{B} ,B)$. By Cauchy-Schwarz, the 
latter is at most $2 \Sigma_{B} \Sigma_{\tilde{B}}$ in norm.\qed 
\noindent We now show that this bound is sharp in the sense that for all 
$S$, $\tilde{S} > 0$, there exist $T$, $B$, $\tilde{B}$
such that (\ref{jointje}) attains equality with 
$\Sigma_{B} = S$, $\Sigma_{\tilde{B}} = \tilde{S}$. 
\subsection{Application: the Beamsplitter as a Joint Measurement}
A \emph{beamsplitter} is a device which takes two beams of light as input. 
A certain fraction of each incident beam is refracted and the rest is
reflected, in such a way that the refracted part of the first beam
coincides with the reflected part of the second and vice versa. 
We will show that the beamsplitter serves as an optimal joint 
unbiased measurement.
\begin{center}
\setlength{\unitlength}{1 cm}
\begin{picture}(8,3.2)
\put(0.1,2.2){\mbox{\footnotesize air}}
\put(0.1,1.7){\mbox{\footnotesize glass}}
\put(0,2){\line(1,0){8}}
\put(4,2){\vector(3,1){3}}
\put(4,2){\vector(1,-1){1.5}}
\put(2.5,0.5){\vector(1,1){1.5}}
\put(1,3){\vector(3,-1){3}}
\end{picture}\\
{\small \fig Beamsplitter.}
\end{center}
In cavity QED, a single mode in the field is described by a Hilbert space
$\CH$ of a harmonic oscillator, with creation and annihilation
operators $a^{\dag}$ and $a$ satisfying $[a,a^{\dag}] = 1$, as well as 
$x = \frac{a + a^{\dag}}{\sqrt{2}}$ and 
$p = \frac{a - a^{\dag}}{\sqrt{2}i}$.
The coherent states $\ket{\alpha} = e^{-|\alpha|^2/2} \sum_{n=0}^{\infty}
\frac{\alpha^n}{\sqrt{n!}} \ket{n}$ are dense in $\CH$, and satisfy
$a\ket{\alpha} = \alpha\ket{\alpha}$.

Quantummechanically, a beamsplitter is described by the unitary operator $U$ on 
$\CH \ten \CH$, given by  
$U = \exp( \theta ( a^{\dag} \ten{a} - a \ten a^{\dag} ))$.
In terms of the coherent vectors, 
we have
$U \ket{\alpha} \ten \ket{\beta} = 
\ket{\alpha \cos(\theta)  + \beta \sin(\theta) } \ten 
\ket{-\alpha\sin(\theta)  + \beta \cos(\theta) }
$. Note that 
$U^{\dag} a\ten\one U = \cos(\theta) a\ten\one   + \sin(\theta) \one\ten a  $
and that
$U^{\dag} \one\ten a U = - \sin(\theta) a\ten\one   + \cos(\theta) \one\ten a $.
(This can be seen by sandwiching both sides between coherent vectors.)
Since the map $Y \mapsto U^{\dag}YU$ respects $+$, $\cdot$ and
${}^{\dag}$, we readily calculate 
\begin{eqnarray*}
U^{\dag} x\ten\one U &=& \cos(\theta) x\ten\one + \sin(\theta) \one\ten x,\\
U^{\dag} x^2\ten\one U &=& \cos^2(\theta) x^2\ten\one + 
2\sin(\theta)\cos(\theta) x\ten x + \sin^2(\theta) \one\ten x^2,\\
U^{\dag} \one\ten p U &=& -\sin(\theta) p\ten\one + \cos(\theta) \one\ten
p,\\ 
U^{\dag} \one\ten p^2 U &=& \sin^2(\theta) p^2\ten\one - 
2\cos(\theta)\sin(\theta) p\ten p + \cos^2(\theta) \one \ten p^2.
\end{eqnarray*}
We are now interested in the map $ \rho \mapsto U \rho \ten \ketbra{0}
U^{\dag}$, from $\CS(\CH)$ to $\CS(\CH) \ten \CS(\CH)$.
In other words, we feed the beamsplitter only one beam of light in a state
$\rho$, the other input being the vacuum. The dual of this is the CP-map
$T : \CB(\CH) \ten \CB(\CH) \rightarrow \CB(\CH)$ defined by 
$T(Y) := id \ten \phi_0 ( U^{\dag}YU )$, with $\phi_0$ the vacuum state
$\phi_0(X) = \bra{0} X \ket{0}$.

Take $B = \cos^{-1}(\theta) x \ten \one$ for instance. Then 
$T(B) = x\bra{0}\one\ket{0} + \tan(\theta) \one \bra{0}x \ket{0} = x$. 
Similarly, with $\tilde{B} = -\sin^{-1}(\theta) \one \ten p$, 
we have $T(\tilde{B}) = p$. Apparently, splitting a beam of light in two parts,
measuring $x\ten\one$ in the first beam and $\one \ten p$ in the second,
and then compensating for the loss of intensity provides a 
\emph{simultaneous unbiased measurement} of $x$ and $p$ in the 
original beam. 
Since $[x,p] = i$, we must\footnote{We neglect the technical complication of
$x$ and $p$ being unbounded operators.
} have $\Sigma_{B} \Sigma_{\tilde{B}} \geq \half$.

We now calculate $\Sigma_{B}$ and $\Sigma_{\tilde{B}}$ explicitly.
From $\bra{0}x^2\ket{0} = \half$, we see that
$T(B^2) = x^2 + \half \tan^2(\theta) \one$. Thus
$\Sigma_{B}^2 = \| (B,B) \| = \half \tan^2 (\theta)$.
Similarly $\Sigma_{\tilde{B}}^2 = \half \tan^{-2}(\theta)$.
We see that $\Sigma_{B}\Sigma_{\tilde{B}} = \half$, so that 
the beamsplitter is indeed an optimal jointly unbiased measurement.

By scaling $B$, optimal joint measurements can be
found for arbitrary values of $\Sigma_{B}$ and $\Sigma_{\tilde{B}}$, 
which shows the bound in
theorem \ref{TJM} to be sharp.
It may therefore be used to evaluate joint measurement procedures.
For example, it was shown in \cite{Luc} that homodyne detection of 
the spontaneous decay 
of a two-level atom constitutes a joint measurement with
$\Sigma \Sigma' = 1.056$, slightly above the bound $\Sigma \Sigma' \geq 1$ 
provided by theorem \ref{TJM}.

The beamsplitter is an optimal joint
measurement in the sense that it minimizes $\Sigma \Sigma'$.
It also performs well with other figures of merit. 
For example, if the quality of joint measurement is judged by the
state-dependent cost  
$R(T):= \mathbf{Var}(B,T^*(\rho)) + \mathbf{Var}(\tilde{B},T^*(\rho))$,
then at least for Gaussian $\rho$, the optimal measurement is again the 
above beamsplitter with $\theta = \pi/4$. (See \cite{Ho}.)

\section{The Heisenberg Principle}\label{Heis}
The Heisenberg Principle may be stated as follows:
\begin{center}
\emph{If all states are left intact, no 
quantum-information can be extracted from a system.}
\end{center}
This alludes to an information transfer from an initial system 
$\CA$ to a final system consisting of two parts: the system 
$\CA$ and an ancilla $\CB$, containing the pointer $B$. 
We thus have an information
transfer $T: \CA \ten \CB \rightarrow \CA$ from $A$ to $\one \ten B$. 

An initial state $\rho \in \CS(\CA)$ gives rise to a final state 
$T^*(\rho) \in \CS(\CA \ten \CB)$. Restricting this final state to the
system $\CA$ (i.e. taking the partial trace over $\CB$)
yields a `residual' state $R^*(\rho) \in \CS(\CA)$, 
whereas taking the partial trace over
$\CA$ yields the final state $Q^*(\rho) \in \CS(\CB)$ of the ancilla.
We define the CP-maps $R: \CA \rightarrow \CA$ by 
$R(A) := T(A \ten \one)$ and $Q: \CB \rightarrow \CA$ by 
$Q(B) := T(\one \ten B)$. The map $R$ describes what happens to $\CA$
if we forget about the ancilla $\CB$, and $Q$ describes the ancilla, 
neglecting the original system $\CA$. 

We wish to find a quantitative version of the Heisenberg principle,
i.e. we want to relate the imperfection of the extracted quantum-information  
to the amount of state disturbance.
\definition The \emph{maximal disturbance} $\Delta$ 
of a map $R : \CA \rightarrow \CA$
is given by 
{$\Delta := \sup \{ \mathrm{ \mbox{$\| R(P) - P \|$} } \,;$
$ \, P \in \CA, P^2 = P^{\dagger} = P \}$}.
\\[0.3 cm] \noindent
The trace distance (or Kolmogorov distance)
$D( \tau , \rho )$ is the maximal difference between the probability 
$\tr(\tau P)$ that an event $P$ occurs in the state $\tau$,
and the probability $\tr( \rho P)$ that it occurs in the state $\rho$,
for the worst case event (projection operator) $P$.
For short,
$D( \tau , \rho ) := 
\sup_{P} \{ | \tr (\tau P) - \tr (\rho P) | \}$.
One may show that 
$D(\tau, \rho) = \half \tr(|\tau - \rho|)$ (see e.g. \cite{NC}).

$\Delta$
is now the worst case distance between the input $\rho$ and 
the output $R^*(\rho)$, i.e.
$\Delta = \sup \{ D( \rho, R^*(\rho) ) ; \rho \in \CS(\CA)\}$.
Indeed, $\sup_{\rho} \{ D( \rho, R^*(\rho) )\} = 
\sup_{\rho, P} \{ \tr (\rho P) - \tr (\rho R(P)) \}$,
which equals
$ \sup_{P} \{ \| R(P) - P \| \} = \Delta$.

\subsection{Heisenberg Principle for Unbiased Information Transfer}
We first turn our attention to unbiased information transfer. 
The imperfection of the information is then captured in the maximal
added variance $\Sigma^2$.

The Heisenberg principle only holds for quantum-information.
Classical observables are contained in the centre 
$\CZ = \{A \in \CA \,; \, [A,X] = 0 \,\,\, \forall X \in \CA \}$, 
whereas quantum observables are not. 
The degree in which an observable $A$ is `quantum'
is given by its distance to the centre 
$d(A,\CZ) = \inf_{Z \in \CZ} \|A - Z\|$.
In the following, we will take the algebra of observables to be
$B(\CH)$ for some Hilbert space $\CH$.
The centre is then simply $\mathbb{C} \one$.

\begin{theorem}\label{HP}
Let $T : B(\CH) \otimes \CB \rightarrow B(\CH)$ be a CP-map, let $B \in \CB$
be Hermitean. Define $A := T(\one \otimes B)$, and 
$\Sigma^2 := \|(\one \ten B,\one \ten B)\|$. 
Furthermore, define 
$\Delta := \sup_{P} \{ \| R(P) - P \| \}$,
with $R$ the restriction of $T$ to $B(\CH) \ten \one$.  
Then 
\begin{equation}\label{bound}
\Sigma \geq d(A,\CZ) \frac{\half -\Delta}{\sqrt{\Delta(1 - \Delta)}}\,.
\end{equation}
This bound is sharp in the sense that for all $\Delta \in [0,\half]$,
there exist $T$ and $A$ for which (\ref{bound}) attains equality. 
\end{theorem}
\proof
For the sharpness, see section \ref{sharp}. As for the bound,
we may assume $\Delta < \half$, since inequality (\ref{bound})
is trivially satisfied otherwise. 
Denote the spectrum of $A$ by $\spec(A)$.
Let $x := \sup(\spec(A))$ and $y := \inf(\spec(A))$, so that 
$d(A,\CZ) = \frac{x - y}{2}$.
Without loss of generality, assume that there exist normalized 
eigenvectors $\psi_x$ and $\psi_y$ satisfying 
$A \psi_x = x \psi_x$ and $A \psi_y = y \psi_y$.
(If this is not the case, choose $x'$ and $y'$ in $\spec(A)$
arbitrarily close to $x$ and $y$, and complete the proof
using approximate eigenvectors.) 
Define $\psi_{-} := \frac{1}{\sqrt{2}} (\psi_x + \psi_y)$,  
$\tilde{B} := \ketbra{\psi_-}$ and
$\tilde{A} := T(\tilde{B} \ten \one)$. 

We thus have
$\|[A,\tilde{B}]\| = d(A,\CZ)$, and furthermore
$\|\tilde{A} - \tilde{B}\| \leq \Delta$, so that
$\|[A, \tilde{A} - \tilde{B}]\| \leq 2 \Delta d(A,\CZ)$.
Then by the triangle inequality 
$\|[T(\one \ten B), T(\tilde{B} \ten \one)]\| = 
\|[A,\tilde{B}] + [A, \tilde{A} - \tilde{B}]\| \geq  d(A,\CZ) (1 - 2\Delta)$,
which brings us in a position to apply theorem \ref{TJM}
to the commuting pointers $\tilde{B} \ten \one$ and $\one \ten B$.
This yields 
\begin{equation}\label{goudvis}
2 \Sigma \sqrt{\|(\tilde{B} \ten \one , \tilde{B} \ten \one)\|} \geq 
d(A,\CZ) (1 - 2 \Delta)\,.
\end{equation}
In order to estimate $\|(\tilde{B} \ten \one, \tilde{B} \ten \one)\|$,
we first prove that $\spec(\tilde{A}) \subseteq [0, \Delta] \cup
[(1 - \Delta),1] $.
Let $a \in \spec(\tilde{A})$. Since $T$ is a contraction and
$0 \leq \tilde{B} \leq 1$, we have $0 \leq a \leq 1$.
Without loss of generality, assume that there
exists a normalized eigenvector $\psi_a$ such that $\tilde{A} \psi_a = a
\psi_a$. (Again, if this is not the case, one may use
approximate eigenvectors.)
Decompose $\psi_a$ over the eigenspaces of $\tilde{B}$, i.e. write
$\psi_a = \chi_1 + \chi_0 $, with 
$\chi_1 \perp \chi_0$, $\tilde{B} \chi_1 = \chi_1$ and 
$\tilde{B} \chi_0 = 0$.
Then $(\tilde{A} - \tilde{B}) \psi_a = (a - 1) \chi_1 + a \chi_0$.
Since $\|\chi_1\|^2 + \|\chi_0\|^2 = 1$, the inequality
$\Delta^2 \geq \|(\tilde{A} - \tilde{B}) \psi_a \|^2 = 
(a-1)^2 \|\chi_1\|^2 + a^2 \|\chi_0\|^2$ implies that
either $|1-a| \leq \Delta$ or $a \leq \Delta$. 
We conclude $\spec(\tilde{A}) \subseteq [0, \Delta] \cup
[(1 - \Delta),1]$, as desired.

This implies  
$\spec(\tilde{A} - \tilde{A}^2) \subseteq [0,\Delta (1 - \Delta )]$.
Since $\tilde{B}^2 = \tilde{B}$, 
we may estimate 
$\|(\tilde{B} \ten \one, \tilde{B} \ten \one )\|
= \| T(\tilde{B} \ten \one) -  T(\tilde{B} \ten \one)^2 \|
= \|\tilde{A} - \tilde{A}^2\| \leq \Delta(1 - \Delta)$.
Combining this with inequality (\ref{goudvis}) yields
$2\Sigma \sqrt{\Delta(1 - \Delta)} \geq d(A,\CZ) (1 - 2\Delta)$,
which was to be demonstrated. \qed


\noindent In the case of no disturbance, $\Delta = 0$, we see that 
$\Sigma \rightarrow \infty$.
No information
transfer from $\CA$ is allowed if all states on $\CA$ are left intact. 
This is Werner's (see \cite{We})
formulation of the Heisenberg principle.
In the opposite case of perfect information transfer, $\Sigma = 0$, 
inequality \ref{bound} shows that $\Delta$ must equal at least one half.
We shall see in section \ref{collapse} that this 
corresponds with a so-called `collapse of the wave
function'. 
These two extreme situations are connected by theorem \ref{HP} in a
continuous fashion, as indicated in the graph below:
\begin{center}
\begin{tabular}{p{8cm}}
\epsfig{file=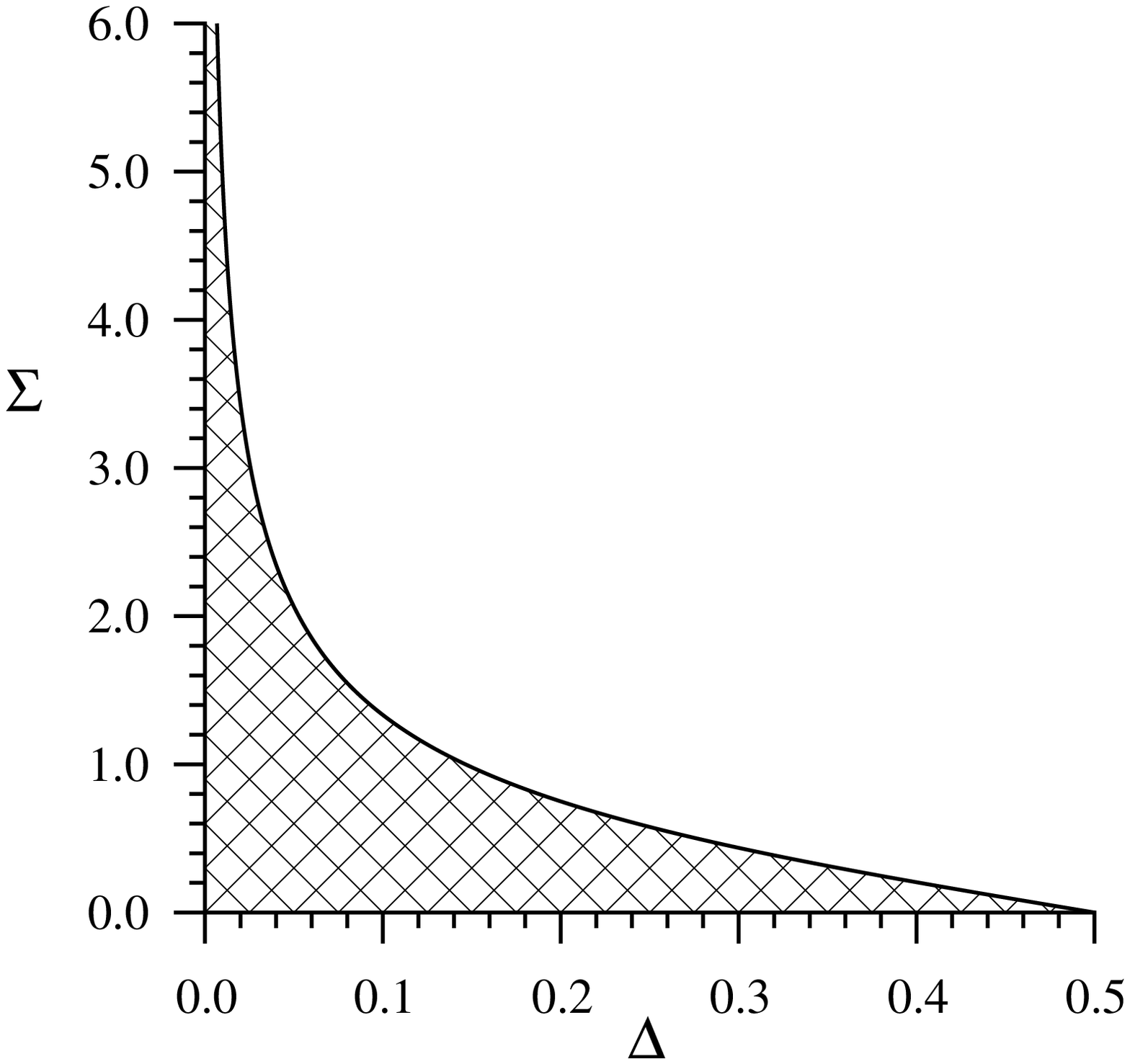, width=6cm}\\[-1 mm]
{\small \fig 
The combinations $(\Delta,\Sigma)$ below the curve 
are forbidden, those above are allowed. 
(With $d(A,\CZ)=1$.)}
\end{tabular}
\end{center}
The upper left corner of the curve illustrates 
the Heisenberg principle,
whereas in the the lower right corner, we can see the collapse of the
wave function.

\subsection{Heisenberg Principle for General Information Transfer}

We now prove a version of the Heisenberg Principle 
for general information transfer. \clearpage

\begin{corollary}\label{hpdel}
Let $T : B(\CH) \otimes \CB \rightarrow B(\CH)$ be a CP-map, let 
$A \in B(\CH)$ and
$B \in \CB$ be Hermitean, $A \notin \CZ = \mathbb{C} \one$. 
Define 
$\Delta := \sup_{P} \{ \| R(P) - P \| \}$,
with $R$ the restriction of $T$ to $B(\CH) \ten \one$.  
Define
$\delta := \sup_{S} \{ \|T(\one \ten \one_{S}(B)) - \one_{S}(A)\| \}$.
Then, for $\delta$ and $\Delta$ in $[0,\half]$, we have
\begin{equation}\label{bound2}
(\half - \delta)^2 + (\half - \Delta)^2 \leq \fourth
\end{equation}
This bound is sharp in the sense that for all 
$\Delta \in [0,\half]$, there exists a $T$
for which (\ref{bound2}) attains equality.
\end{corollary}
\begin{center}
\begin{tabular}{p{8cm}}
\epsfig{file=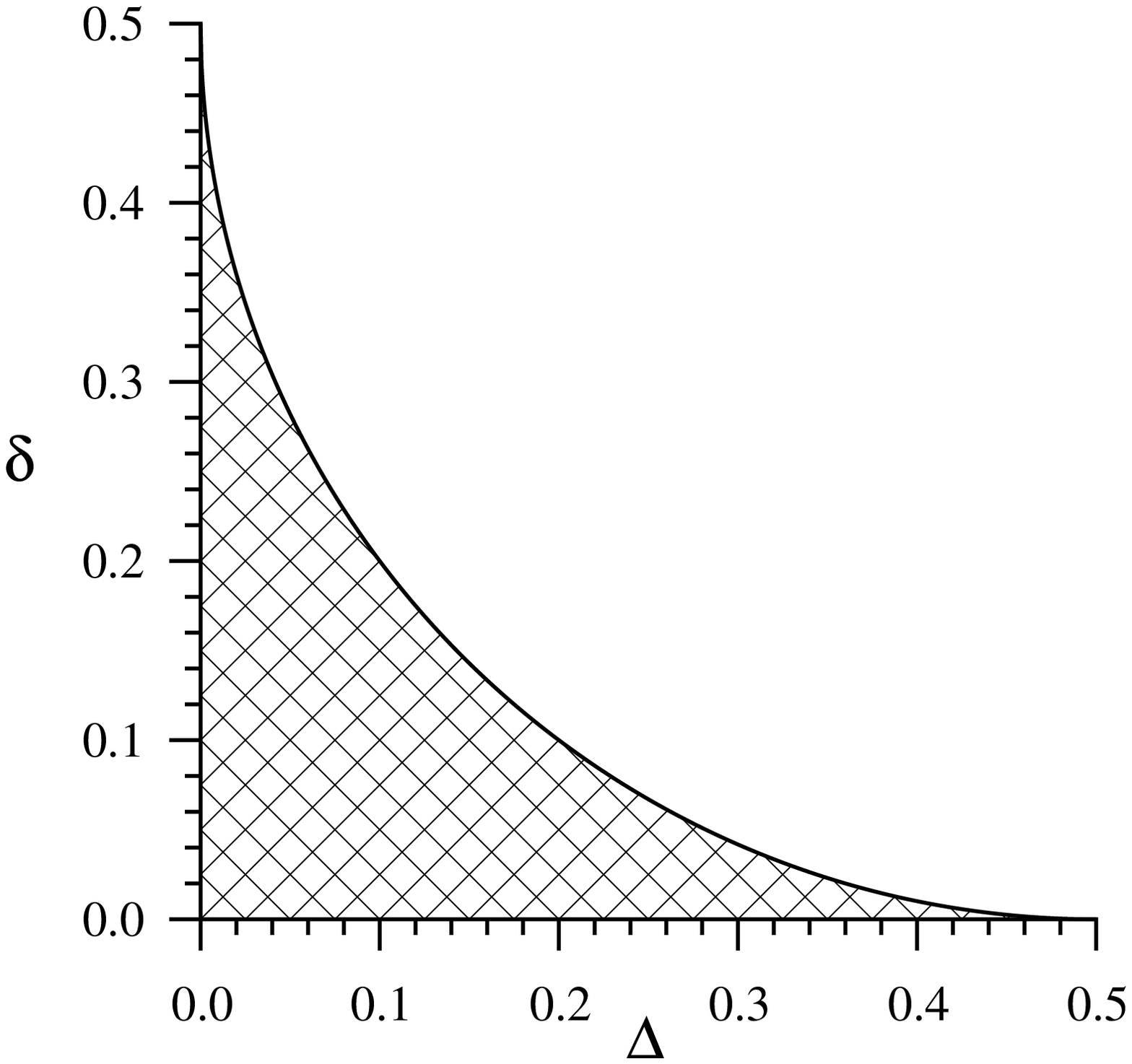, width=6cm}\\[-1mm]
{\small \fig 
The combinations $(\Delta,\delta)$ below the curve 
are forbidden, those above are allowed.}
\end{tabular}
\end{center}
\proof
Choose a nontrivial subset $S$
of $\spec(A)$ and put $P := \one \ten \one_{S}(B)$. 
Since $ \| T(P) - \one_{S}(A) \| \leq \delta$
and $\spec(\one_{S}(A)) = \{0,1\}$, we have
$\spec(T(P)) \subseteq 
[0,\delta] \cup [1-\delta, 1]$ (cf. the proof of theorem 
\ref{HP}).
Thus $\Sigma^2 = \|T(P) - T(P)^2 \|
\leq \delta(1-\delta)$.
Similarly, $d(T(P), \CZ) \geq \half- \delta$
since $\spec(T(P))$ contains points in both
$[0,\delta]$ and $[1 - \delta,1]$.
Apply theorem \ref{HP} to the pointer $P$
to obtain 
$
\sqrt{\delta(1-\delta)} \geq
(\half - \delta)(\half - \Delta)/\sqrt{\Delta(1-\Delta)}
$,
or equivalently 
$(\half - \delta)^2 + (\half - \Delta)^2 \leq \fourth$.
For sharpness, see section \ref{sharp}. 
\qed
A measurement which does not disturb any state ($\Delta = 0$)
cannot yield information ($\delta \geq \half$).
This is the Heisenberg principle.
On the other hand, perfect information ($\delta = 0$)
implies full disturbance ($\Delta \geq \half$), corresponding
to the collapse of the wave function.
Both extremes are connected in a continuous fashion, 
as depicted above.
 
\subsection{Application: Resonance Fluorescence} \label{Resflu}
Corollary \ref{hpdel} may be used to determine the minimum amount of 
disturbance if the quality
of the measurement is known. Alternatively, if the system is only
mildly disturbed, one may  
find a bound on the attainable measurement quality. 
Let us concentrate on the latter option.

We investigate the radiation emission of a laser-driven two-level atom.
The emitted EM-radiation yields information on the atom.
A two-level atom (i.e. a qubit) only has three independent
observables: $\s_x$, $\s_y$ and $\s_z$. 
There are various ways to probe the EM field: photon counting, 
homodyne detection, heterodyne 
detection, etcetera. 
For a strong ($\OO \gg 1$) resonant 
($\omega_{\mathrm{laser}} = \omega_{\mathrm{atom}}$)
laser, we will use corollary \ref{hpdel} to prove that \emph{any} 
EM-measurement of $\s_x$, $\s_y$ or 
$\s_z$ will have a 
measurement infidelity of at least
$$
\delta \geq 
\half - \half \sqrt{1 - e^{-\frac{3}{2}\lambda^{2} t}}\,,
$$
with $\lambda$ the coupling constant.
For a measurement with two outcomes,
$\delta$ is the maximal probability of getting the
wrong outcome.
\\[-0.5cm]
\begin{center}
\begin{tabular}{p{8cm}}
\epsfig{file=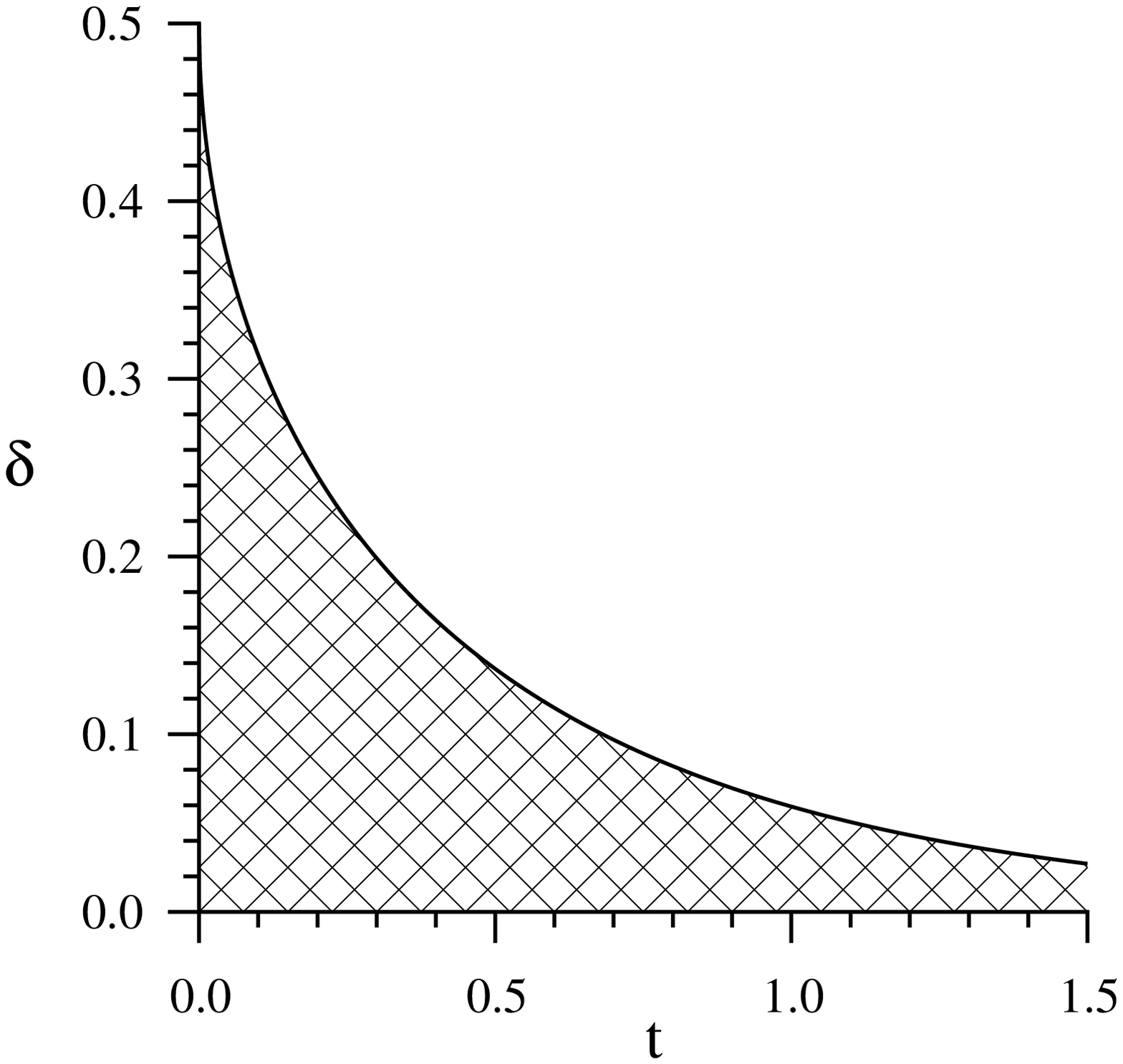, width=6cm}\\[-1 mm]
{\small \fig Lower bound on $\delta$ in terms of $t$ 
(in units of $\lambda^{-2}$).}
\end{tabular}
\end{center}

\subsubsection{Unitary Evolution on the Closed System} 
The atom is modelled by the Hilbert-space $\mathbb{C}^2$ (only two
energy-levels are deemed relevant). In the field, we discern a 
\emph{forward} and a \emph{side} channel, each described by a 
bosonic Fock-space $\CF$.  
The laser is put on the forward channel, which is thus initially in the 
state $\phi_\OO$, the coherent state with frequency $\omega$ and strength
$\OO$. 
(The field strength is parametrized by the frequency of the induced Rabi-oscillations).
The side channel starts in the vacuum state $\phi_0$.   
The state at time $t$ is then given by 
$$
T_{t}^* (\rho) =  U( t ) (\rho \ten \phi_{\OO} \ten \phi_0 )U^{\dag}(t)\,,
$$
with time evolution 
$$
\frac{d}{dt}  U_t = -i(H_S + H_F + \lambda H_I) U_t \,.
$$
$H_S \in B(\mathbb{C}^2)$ is the Hamiltonian of the two-level atom,
$H_F \in B(\CF \otimes \CF)$ that of the field and 
$\lambda H_I \in B(\mathbb{C}^2) \ten B(\CF \ten \CF)$ is the 
interaction-Hamiltonian.
Define the interaction-picture time evolution by 
$$
\hat{T}^*_t (\rho):= U_1 (t)^{\dag}U_2 (t)^{\dag} T^*_t (\rho) 
U_2 (t) U_1 (t)\,,
$$
where
$U_1 (t) := e^{-iH_S t}$ and
$U_2 (t) := e^{- i H_F t}$
form
the `unperturbed' time evolution. 

We now investigate $\hat{T}_t$ instead of $T_t$.
Indeed, we are looking for a bound on the measurement infidelity 
$\delta = \sup_{S}\{ \|T(\one \ten \one_S(B)) - \one_S(A)\| \}$ of $T$.
Yet if $\hat{B} := U_2^{\dag}B U_2$, then 
$\hat{T}(\one_S (\hat{B})) = T(\one_S(B))$, so that
$\hat{\delta} = 
\sup_{S}\{ \|\hat{T}(\one \ten \one_S(\hat{B})) - \one_S(A)\| \}
= \delta$.
If we find the interaction-picture disturbance $\hat{\Delta}$,
corollary \ref{hpdel} will yield a bound on 
$\hat{\delta}$, and thus on $\delta$.

In the weak coupling limit $\lambda \downarrow 0$, $\hat{T}_t$ is given by 
$\hat{T}_{t}^* (\rho) = 
\hat{U}( t/\lambda^2 ) 
(\rho \ten \phi_{\OO} \ten \phi_0 )
\hat{U}^{\dag}(t/\lambda^2)$,
where the evolution of the unitary cocycle
$t \mapsto \hat{U}_t$ is described (see \cite{AFL}) by a 
Quantum Stochastic Differential Equation 
or QSDE. Explicitly calculating the maximal added variances $\Sigma^2$ by solving the
QSDE is in general rather nontrivial, if possible at all. ( See \cite{Luc} for the 
case of spontaneous decay, i.e. $\OO = 0$, with the map $\hat{T}_t$ 
restricted to the commutative algebra of homodyne 
measurement results.)    

\subsubsection{Master Equation for the Open System}

Fortunately, in contrast to the somewhat complicated time evolution
$\hat{T}_t$ of the 
combined system, the evolution \emph{restricted to the 
two-level system} is both well-known and uncomplicated.
If we use $\lambda^{-2}$ as a unit of time, then
the restricted evolution 
$\hat{R}^*_{t}(\rho) := \tr_{\CF \ten \CF} \hat{T}^{*}_{t} (\rho)$
of the two-level system
is known (see \cite{Bo}) to satisfy the Master equation 
\begin{equation} \label{bitrev}
\frac{d}{dt} \hat{R}_{t}^* (\rho) = L(\hat{R}_{t}^* (\rho))\,,
\end{equation}
with the Liouvillian
$
L(\rho) := 
\half i \OO [e^{-i(\omega - E) t} V + e^{i(\omega - E)t} V^{\dag}, \rho] -
\half \{ V^\dag V , \rho  \} + V \rho V^{\dag}
$. In this expression,
$E$ is the energy-spacing of the two-level atom 
and $V^\dag = \s_+$, $V = \s_-$ are its raising and lowering operators.
In the case $\omega = E$ of \emph{resonance fluorescence}, 
we obtain  
$$
L(\rho) = 
\half i \OO [ V + V^{\dag}, \rho] -
\half \{ V^\dag V , \rho  \} 
+ V \rho V^{\dag}\,.
$$
If we parametrize a state by its Bloch-vector 
$\hat{R}^*_t (\rho) = \half(\one + x \s_x + y \s_y + z \s_z)$, 
then equation \ref{bitrev}
is simply the following differential equation on $\mathbb{R}^3$:
$$
\frac{d}{dt} \pmatrix{x \cr y \cr z} =
\pmatrix{	-\half & \,0& \, 0 \cr 
			0 &-\half& \, \OO \cr
			0 & -\OO & -1}
\pmatrix{x \cr y \cr z}
- \pmatrix{0 \cr 0 \cr 1}			
$$
This can be solved explicitly.
For $\OO \gg 1$, the solution approaches
$$\pmatrix{x \cr y \cr z} = 
\pmatrix{e^{-\half t} & 0 & 0 \cr
		0 & \, e^{-\frac{3}{4} t} \cos(\OO t) & e^{-\frac{3}{4} t}	\sin( \OO t)	\cr
		0 & - e^{-\frac{3}{4} t} \sin(\OO t)& e^{-\frac{3}{4} t} \cos(\OO t)}
\pmatrix{x_0 \cr y_0 \cr z_0}\,.
$$
If we move to the interaction picture 
once more to counteract the Rabi-oscillations, 
i.e. with 
$U_1(t) = e^{\frac{i}{2} \OO t \s_x}$ and $U_2 = \one$,
we see that the time evolution is transformed to
$$
\pmatrix{x \cr y \cr z} = 
\pmatrix{e^{-\half t} & 0 & 0 \cr
		0 & \, e^{-\frac{3}{4} t}  & 0	\cr
		0 &  0 & e^{-\frac{3}{4} t} }
\pmatrix{x_0 \cr y_0 \cr z_0}\,.
$$
Since the trace distance $D( \rho , \tau )$ is exactly half the Euclidean distance between the
Bloch vectors of $\rho$ and $\tau$, (see \cite{NC}), we see that $\Delta = \half(1- e^{-\frac{3}{4}t})$.
For any measurement of $\s_x$, $\s_y$ or $\s_z$, we therefore have
$
\delta \geq \half - \half \sqrt{1 - e^{-\frac{3}{2} t}}
$ by corollary \ref{hpdel}
(remember that $t$ is in units of $\lambda^{-2}$).

\section{Collapse of the Wave function}\label{collapse}
The `collapse of the wave function' may be seen as the flip side
of the Heisenberg principle. It states that if information is
extracted from a system, then its states undergo a very specific
kind of perturbation, called decoherence.

\subsection{Collapse for Unbiased Information Transfer}
We start out by investigating unbiased Information Transfer. 
We prove a sharp upper bound on the amount of remaining coherence
in terms of the measurement quality. 
\begin{theorem} \label{offdiag}
Let $T : \CA\ten\CB \rightarrow \CA$ be a CP-map. Let $B \in \CB$ 
be Hermitian, and define $A := T(\one \ten B)$.
Suppose that $\psi_x$ and $\psi_y$ are eigenvectors of $A$ with 
different eigenvalues 
$x$ and $y$ respectively.
Define $R: \CA \rightarrow \CA$ to be the restriction of 
$T$ to $\CA \ten \one$, and put $\Sigma^2 := \|(B,B)\|$.
Then for all $\alpha,\beta \in \mathbb{C}$ with $|\alpha|^2 + |\beta|^2 = 1$,
we have
\begin{equation}\label{karper}
D \Big(
R^* \big( \ketbra{\alpha \psi_x + \beta \psi_y} \big) , 
R^*\big( |\alpha|^2 \ketbra{\psi_x} + |\beta|^2 \ketbra{\psi_y} \big) 
\Big) \leq
\frac{ \Sigma / |x-y| }
{\sqrt{1 + 4 \left( \Sigma / |x-y| \right)^2}} \,.
\end{equation}  
This bound is sharp in the sense that for all values of $\Sigma/|x-y|$,
there exist $T$, $\psi_x$, $\psi_y$, $\alpha$ and $\beta$ for which
(\ref{karper}) attains equality.
\end{theorem}
\proof 
The l.h.s. of (\ref{karper}) equals
$ 
\sup\{ \bar{\alpha}\beta \inp{\psi_x}{R(P) \psi_y} + \mathrm{\textbf{c.c.}}
\, | \,
P \in \CA,  P^2=P^{\dag}=P \}   
$. Furthermore, $ 2|\alpha||\beta| \leq 1$,
so that it suffices to bound the `coherence' 
$\inp{\psi_x}{R(P) \psi_y}$ on all projections $P$. 
Now  
$(x-y) \inp{\psi_x}{R(P) \psi_y} =
\inp{\psi_x}{[A,R(P)] \psi_y}$, and
$[A,R(P)] = 
(P \ten \one, \one \ten B) - 
(\one \ten B, P \ten \one)$. Thus
\begin{equation} \label{garbok}
(x-y) \inp{\psi_x}{R(P) \psi_y} =
\inp{\psi_x}{(P \ten \one, \one \ten B) \psi_y} - 
\inp{\psi_x}{(\one \ten B, P \ten \one) \psi_y}\,,
\end{equation}
and we will bound these last two terms.
In the notation of lemma \ref{causch}, we have
$\| g(\one \ten B) \| = \Sigma$. Therefore
$
\inp{\psi_x}{(P \ten \one, \one \ten B) \psi_y}
=
\inp{g(P \ten \one) \psi_x}{g(\one \ten B)\psi_y}
\leq
\|g(P \ten \one) \psi_x\| \|g(\one \ten B)\psi_y\| 
\leq
\Sigma \sqrt{\inp{\psi_x}{(P\ten\one , P\ten\one) \psi_x}}\,.
$
We will bound 
$
\inp{\psi_x}{ (T(P^2 \ten \one) - T(P\ten \one)^2) \psi_x } =
\inp{\psi_x}{ (R(P) - R(P)^2) \psi_x }
$
in terms of the coherence.
For brevity, denote $X_{xx'} := \inp{\psi_x}{X \psi_{x'}}$.
Since $\psi_x \perp \psi_y$,  we have 
$
(R(P)^2)_{xx} \geq |R(P)_{xx}|^2 + |R(P)_{xy}|^2
$, so that 
$ 
(R(P) - R(P)^2)_{xx}
\leq
R(P)_{xx}(1- R(P)_{xx}) - |R(P)_{xy}|^2
$.
Since $x(1-x) \leq \fourth$ for all $x \in \mathbb{R}$,
this is at most
$
\fourth - |R(P)_{xy}|^2
$. 
All in all, we have obtained
$
(P \ten \one, \one \ten B)_{xy} \leq 
\Sigma \sqrt{ \fourth - |R(P)_{xy}|^2 }
$, and of course the same for $x \leftrightarrow y$.
Plugging these into equation \ref{garbok} yields
$
|x-y| |R(P)_{xy}| \leq 
2 \Sigma  \sqrt{\fourth - |R(P)_{xy}|^2}\,,
$
or equivalently 
$
|R(P)_{xy}| \leq 
\frac{ \Sigma / |x-y| }
{\sqrt{1 + 4 \left( \Sigma / |x-y| \right)^2}} 
$, which was to be proven. For sharpness, see section \ref{sharp}.
\qed
\noindent Consider the ideal case of perfect ($\Sigma = 0$) 
information transfer.
Suppose that the system
$\CA$ is initially 
in the coherent state $\ketbra{\alpha \psi_x + \beta \psi_y}$.
Then 
theorem \ref{offdiag} says that, after the information 
transfer 
to the ancilla $\CB$, the system $\CA$
cannot be distinguished from one that started out
in the `incoherent' 
state $|\alpha|^2 \ketbra{\psi_x} + |\beta|^2 \ketbra{\psi_y}$.
As far as the behaviour of $\CA$ 
is concerned, it is therefore 
completely harmless to assume that
a collapse 
$
\ketbra{\alpha \psi_x + \beta \psi_y}
\mapsto
|\alpha|^2 \ketbra{\psi_x} + |\beta|^2 \ketbra{\psi_y}
$
has occurred at the start of the procedure.

Now consider the other extreme of a measurement which leaves all states intact,
i.e. $R^*(\rho) = \rho$ for all $\rho$. 
Then there exist states for which the l.h.s. of
equation (\ref{karper}) equals $\half$, forcing
$\Sigma \rightarrow \infty$; no information can be obtained. 
This is Werner's formulation of the Heisenberg principle.

Theorem \ref{offdiag} thus unifies the Heisenberg principle
and the collapse of the wave function. 
For $\Sigma = 0$ we have a full decoherence,
whereas if all states are left intact, we have $\Sigma \rightarrow \infty$. 
For all intermediate cases,
the bound \ref{karper}
on the remaining coherence
is an increasing function of 
$\Sigma / |x - y|$.\\[-3 mm] 
\begin{center}
\begin{tabular}{p{8cm}}
\epsfig{file=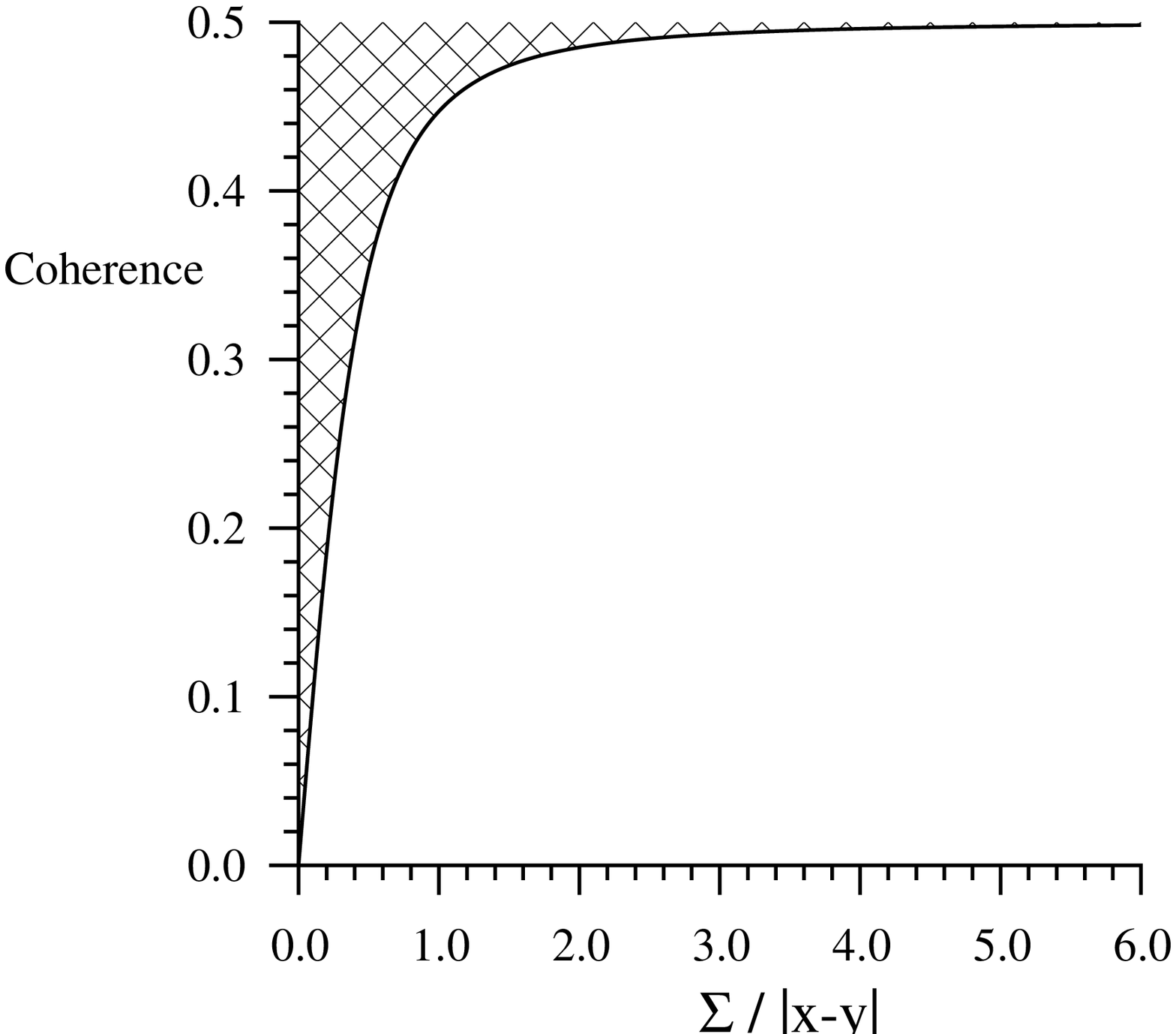, width=6cm}\\
{\small \fig Bound on the coherence as a function of $\Sigma/|x-y|$.
All points above this curve are forbidden, all points below
are allowed.}
\end{tabular}
\end{center}
This agrees with physical intuition: 
decoherence between 
$\psi_x$ and $\psi_y$ is expected to occur in case the information
transfer is able to distinguish between the two. 
This is the case if the variance is small w.r.t. the differences in mean.

\subsection{Application: Perfect Qubit Measurement}
In section \ref{trans}, we have encountered 
the von Neumann Qubit measurement.
Now consider any perfect measurement $T$ of $\s_z$
with pointer $\one \ten (\delta_+ - \delta_-)$ which
leaves $\up$ and $\down$ in place, i.e.
$R^*(\up) = \up$ and $R^*(\down) = \down$.
(Such a measurement is called nondestructive.) 
Theorem \ref{offdiag} then reads 
$R^*(\ketbra{\alpha \uparrow + \beta \downarrow}) = 
|\alpha|^2 \ket{\!\uparrow\,}\bra{\,\uparrow\!} + 
|\beta|^2 \ket{\!\downarrow\,}\bra{\,\downarrow\!}$, 
illustrated below.
\begin{center}
\begin{tabular}{c}
\epsfig{file=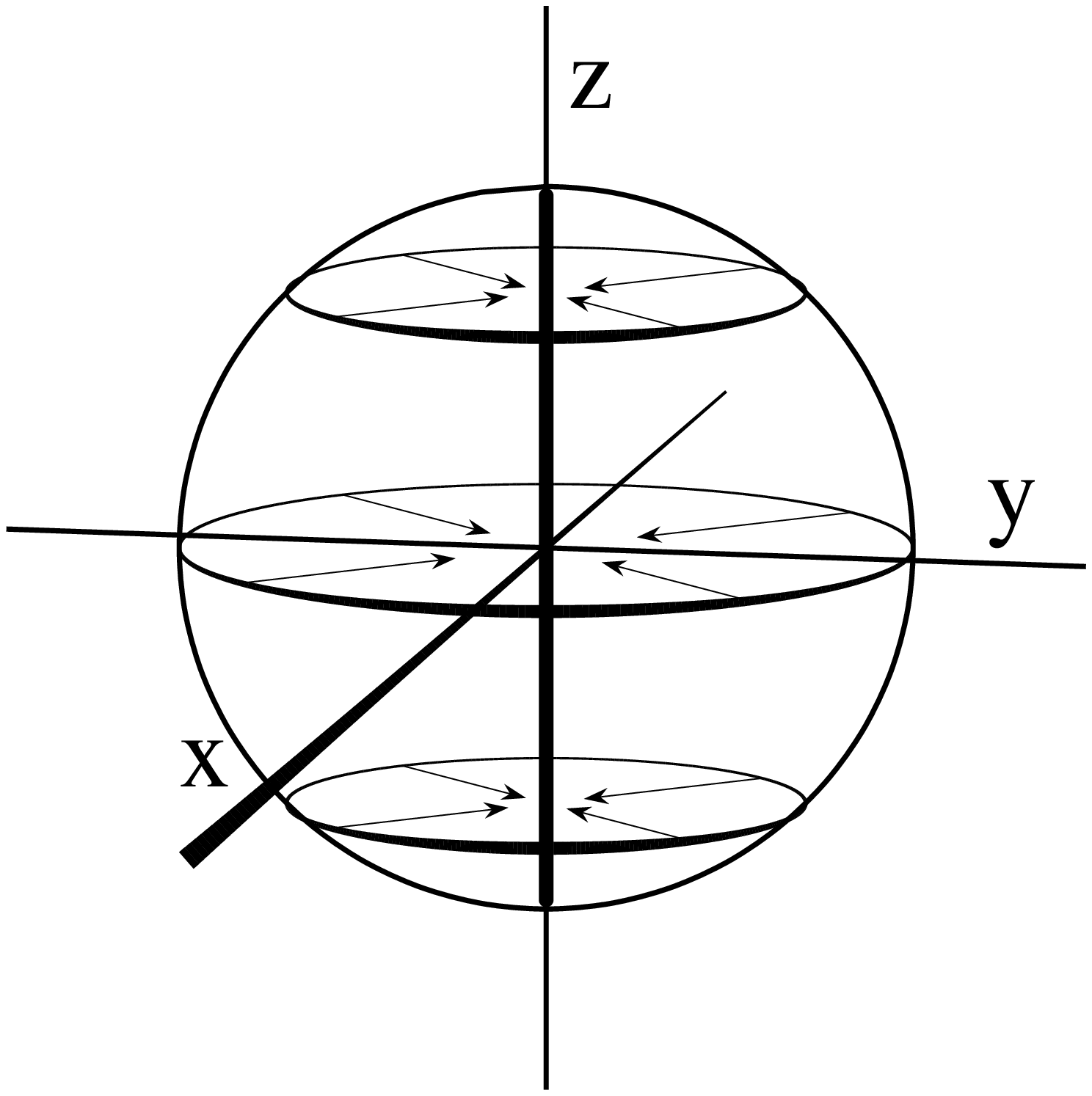, width=6cm}\\
{\small \fig Collapse on the Bloch-sphere for perfect measurement.}\\ 
\end{tabular}
\end{center}
Incidentally, the trace distance between the centre of the Bloch sphere and
its surface is $\half$, so that we read off
$\Delta = \sup \{ D(R^*(\rho),\rho) ; \rho \in \CS(M_2) \} = \half$.
This was predicted by theorem \ref{HP}.

\subsection{Collapse of the Wave function for General Measurement }

We will prove a sharp bound on the remaining coherence 
in general information transfer.
For technical convenience, we will focus attention on 
nondestructive measurements.
A measurement of $A$ is called `nondestructive' 
(or `conserving' or `quantum nondemolition')
if it leaves the eigenstates of $A$ intact,
so that repetition of the measurement will yield the same result.
For example, the measurement in section \ref{Neuqubit}
is nondestructive, the one in section \ref{Resflu} is destructive.
Restriction to nondestructive measurements is quite common  
in quantum measurement theory (see \cite{Per}). 

\begin{corollary}\label{gencol}
Let $T : B(\CH) \otimes \CB \rightarrow B(\CH)$ be a CP-map, let 
$A \in B(\CH)$ and
$B \in \CB$ be Hermitean and 
let $\{ \psi_{i}\}$ be an orthogonal basis of eigenvectors of $A$,
with eigenvalues $a_i$.
Define the measurement infidelity
$\delta := \sup_{S} \{ \|T(\one \ten \one_{S}(B)) - \one_{S}(A)\| \}$.  
Suppose that $T$ is nondestructive, i.e.
$R^*(\ketbra{\psi_{i}}) = \ketbra{\psi_{i}}$ for all $\psi_{i}$,
with $R$ the restriction of $T$ to $B(\CH) \ten \one$.
Then if $\delta \in [0 , \half]$, and $a_i \neq a_j$,
\begin{equation}\label{bound3}
D\Big( 
R^* \big(\ket{\alpha \psi_{i} + \beta \psi_{j} \,} \bra{ \, \alpha \psi_{i} +
\beta \psi_{j} } \big)  , 
\big( |\alpha|^2 \ketbra{\psi_{i}} + 
|\beta|^2 \ketbra{\psi_{j}} \big) 
\Big) 
\leq \sqrt{\delta(1-\delta)}\,.
\end{equation}
This bound is sharp in the sense that for all $\delta \in [0,\half]$,
there exist $T$, $\psi_{i}$, $\psi_{j}$, $\alpha$ and $\beta$ 
for which (\ref{bound3}) attains equality.
\end{corollary}
\begin{center}
\begin{tabular}{p{8cm}}
\epsfig{file=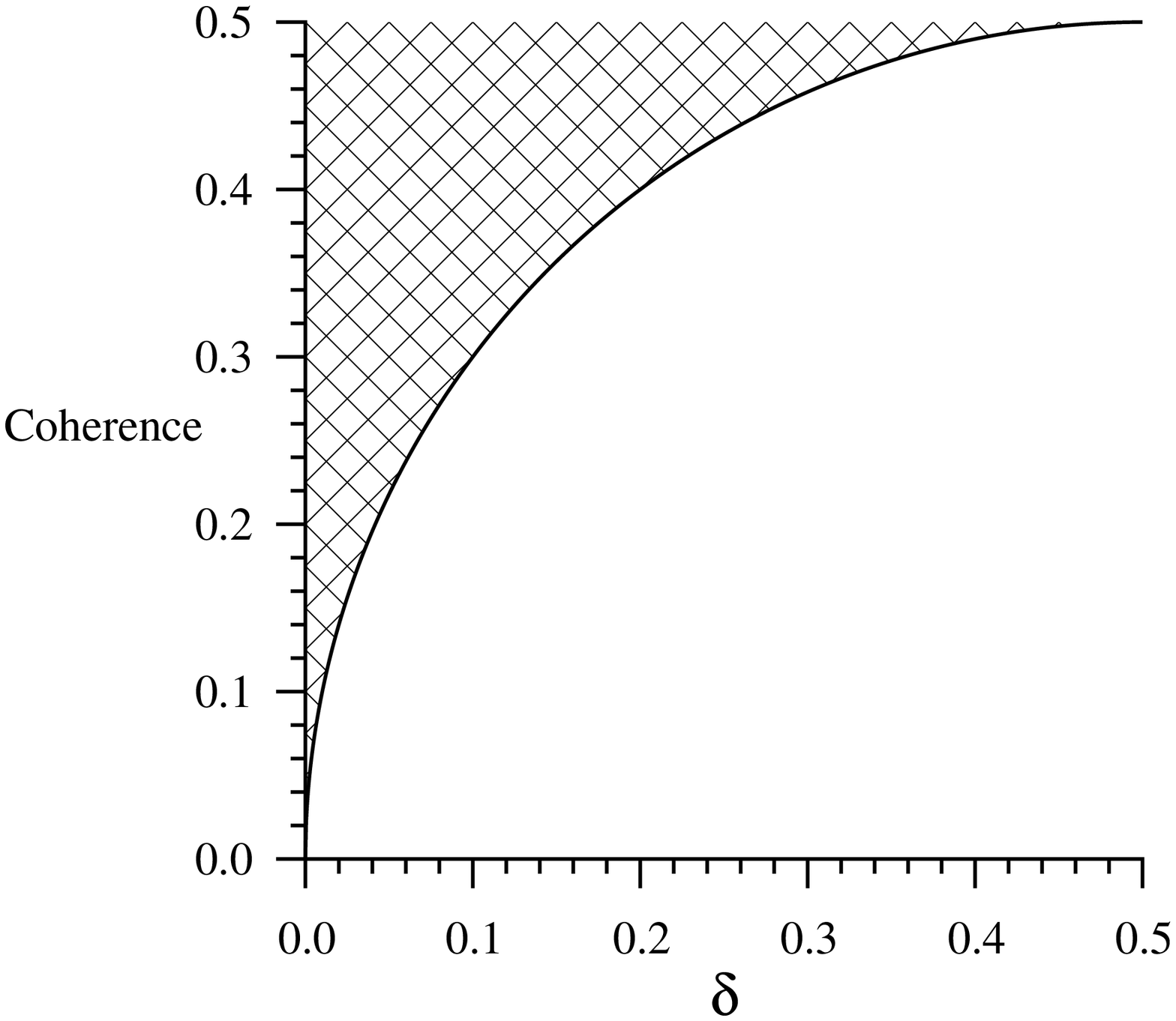, width=6cm} \\[-1mm]
{\small \quad \fig \label{figr}
Bound on the coherence in terms of $\delta$.
All points above the curve are forbidden,
all points below are allowed.}
\end{tabular}
\end{center}
\proof
For sharpness, see section \ref{sharp}. 
Choose a set $S$ such that $a_i \in S$ and $a_j \notin S$.
$T$ is an unbiased measurement of $T(\one \ten \one_{S}(B))$
with pointer $\one_{S}(B)$ and maximal added variance
$\Sigma^2 \leq \delta(1 - \delta)$ (cf. the proof of corollary
\ref{hpdel}).
We will prove that $\psi_{i}$ and $\psi_{j}$ are
eigenvectors of $T(\one \ten \one_{S}(B))$ with  
eigenvalues $x$ and $y$ which differ at least $1 - 2\delta$.

Define $P_{i} := \ketbra{\psi_{i}}$. Since $T$ is nondestructive, 
we have
$\inp{\psi_{j}}{ R(P_{i}) \psi_{j}} = 
\inp{\psi_{j}}{ P_{i} \, \psi_{j}}$
for all $j$.
Apparently, $R(P_{i})$ has only one nonzero diagonal element, a $1$ 
at position $(i,i)$. Since $R(P_{i}) \geq 0$, this implies
$R(P_{i}) = P_{i}$. Then 
$(P_{i} \ten \one , P_{i} \ten \one ) = 0$,
so that by Cauchy-Schwarz
$(P_{i} \ten \one , \one \ten \one_{S}(B)) = 0$.
Since $P_{i} = T( P_{i} \ten \one )$, we have
$
[T(\one \ten \one_{S}(B)) , P_{i} ] =
(P_{i} \ten \one , \one \ten \one_{S}(B)) -
(\one \ten \one_{S}(B) , P_{i} \ten \one)
= 0
$. Therefore $\psi_{i}$ is an eigenvector of
$T(\one \ten \one_{S}(B))$, with eigenvalue $x$, say. 
By a similar reasoning, $\psi_{j}$ is also an eigenvector,
denote its eigenvalue by $y$.

Since $\| T(\one \ten \one_{S}(B)) - \one_S (A) \| \leq \delta$,
we have in particular 
$\| (T(\one \ten \one_{S}(B)) - \one_S (A)) \psi_{i} \| = 
|x-1| \leq \delta$
and
$\| (T(\one \ten \one_{S}(B)) - \one_S (A)) \psi_{j} \| 
= |y| \leq \delta$, so that $|x - y| \geq 1 - 2\delta$.
We can now apply Theorem \ref{offdiag}. On the l.h.s.
of the bound (\ref{karper}), we may substitute
$R^*(|\alpha|^2 \ketbra{\psi_{i}} + 
|\beta|^2 \ketbra{\psi_{j}}) = 
|\alpha|^2 \ketbra{\psi_{i}} + 
|\beta|^2 \ketbra{\psi_{j}}$ 
on account of $T$ being nondestructive.
On the r.h.s., 
we substitute $\Sigma = \sqrt{\delta(1-\delta)}$ 
and $|x - y| = (1 - 2\delta)$. 
Strikingly enough, this yields the bound
$\big( \sqrt{\delta(1-\delta)} / (1-2\delta) \big)   / 
\sqrt{1 + 4 
\big( \delta(1-\delta) / (1 - 2\delta)^2 \big)} = 
\sqrt{\delta(1-\delta)}$. \qed
For perfect measurement ($\delta = 0$), this yields
$R^* (\ket{\alpha \psi_{i} + \beta \psi_{j} \,} \bra{ \, \alpha \psi_{i} +
\beta \psi_{j} } )  =  |\alpha|^2 \ketbra{\psi_{i}} + 
|\beta|^2 \ketbra{\psi_{j}}$; all coherence between $\psi_i$ and $\psi_j$
must vanish. 
This collapse of the wave function is illustrated in the lower left corner of fig. \ref{figr}.
On the other hand, if all states are left intact so that 
$R^* = \mathrm{\it Id}$, then we must have $\delta = \half$;
no information can be gained.
This is illustrated in the upper right corner of fig.\ref{figr}.
Corollary \ref{gencol} is a unified description of the Heisenberg principle
and the collapse of the wave function.

\subsection{Application: Nondestructive Qubit-Measurement}
In quantum information theory, a
$\s_z$-measurement is often taken to yield output $+1$
or $-1$, according to whether the input was 
$\ket{\!\uparrow\,}$ or $\ket{\!\downarrow\,}$.
It is nondestructive if it leaves the states
$\ket{\!\uparrow\,}$ and $\ket{\!\downarrow\,}$ intact,
yet it is only unbiased if it is perfect. 
Corollary \ref{gencol} shows that in the nondestructive case, 
the Bloch-sphere 
collapses to the cigar-shaped region depicted below:  
\begin{center}
\begin{tabular}{c}
\epsfig{file=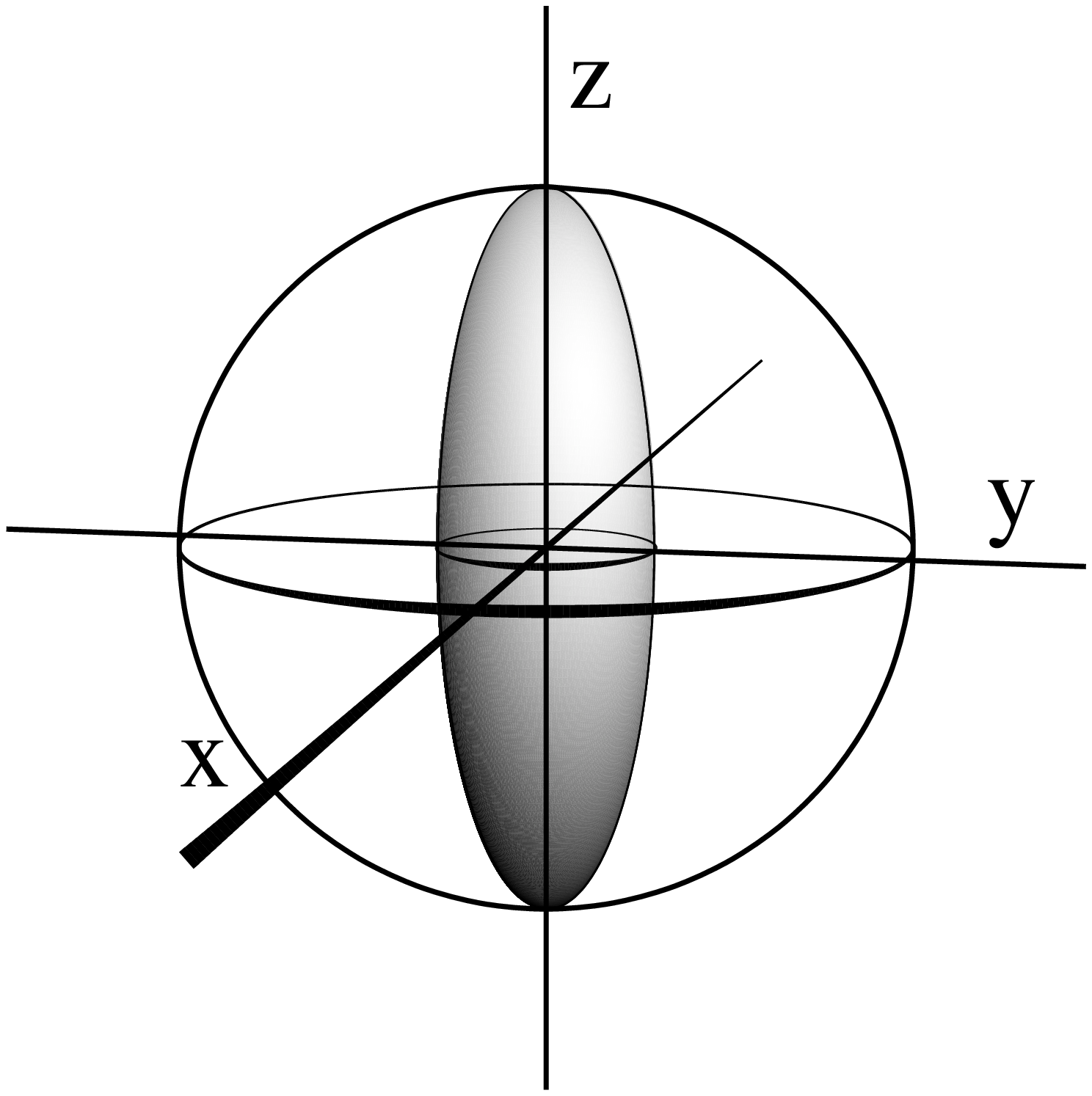, width=6cm}\\
{\small \fig Collapse on the Bloch-sphere with $\delta = 0.01$}.
\end{tabular}
\end{center}
Current single-qubit readout technology is just now moving into 
the regime where the bound (\ref{bound3}) becomes significant. 
in \cite{Lu}, a nondestructive measurement
of a SQUID-qubit was described, with experimentally determined 
measurement infidelity $\delta = 0.13$.
The bound then equals 0.336.

\subsection{Sharpness of the Bounds}\label{sharp}

We have yet to prove sharpness of all bounds.
Let  $V_+ := \pmatrix{\sqrt{1-p} & 0 \cr 0 & \sqrt{p}}$,
$V_- := \pmatrix{\sqrt{p} & 0 \cr 0 & \sqrt{1-p}}$, and define 
$T :  M_2 \ten \CC(\OO) \rightarrow M_2$ by
$T(X \ten f) := f(+1) V_+ X V_+ + f(-1) V_- X V_-$.
For $p=0$, this is the von Neumann-measurement. 
As a measurement of $\s_z$
with pointer $B:= (\delta_+ - \delta_-)/(1-2p)$,
we have
$\delta = p$.
This yields bounds on the disturbance and on the coherence.
Corollary \ref{hpdel} and theorem \ref{HP} yield
$\Delta \geq \half - \sqrt{p(1-p)}$,
corollary \ref{gencol} and theorem \ref{offdiag} yield  
$
D\big(
R^* \big(\ket{\alpha \uparrow + \beta \downarrow \,} \bra{\alpha \uparrow +
\beta \downarrow \!} \big)   
,$
$
\big( |\alpha|^2 \ket{\! \uparrow \,}\bra{\, \uparrow \!} + 
|\beta|^2 \ket{\! \downarrow \,}\bra{\, \downarrow \!} \big) 
\big)
\leq \sqrt{p(1-p)}
$.
We now explicitly calculate the restriction of $T$ to $M_2$,
and find 
$$R^*(\rho) = \pmatrix{\rho_{11} & 2\sqrt{p(1-p)} \rho_{12} 
\cr 2\sqrt{p(1-p)} \rho_{21} & \rho_{22}}\,.$$
The maximal remaining coherence occurs for $\alpha = \beta = 1/\sqrt{2}$,
for which it equals $\sqrt{p(1-p)}$. 
The maximal disturbance equals $\Delta = \half - \sqrt{p(1-p)}$.
This shows all bounds to be sharp.

\section{Conclusion}
Our investigation of joint measurement, the Heisenberg principle and
decoherence has yielded the following results.
\begin{itemize}
\item[I]
Theorem \ref{TJM} provides a sharp, state independent bound on 
the performance of unbiased joint measurement of noncommuting observables. 
In the case of perfect ($\Sigma=0$)
measurement of one observable, it implies that 
no information whatsoever 
($\Sigma' = \infty$) can be gained on the other. 
\clearpage
\item[II]
Theorem \ref{HP} (for unbiased information transfer)
and corollary \ref{hpdel} (for general information transfer) 
provide a sharp, state independent bound on the 
performance of a measurement in terms of the 
maximal disturbance that it causes.
In the case of zero disturbance, when all states are left intact,
it follows that no information can be obtained.
This is the Heisenberg principle.
\item[III]
Theorem \ref{offdiag} (for unbiased information transfer) and 
corollary \ref{gencol} (for general information transfer) 
provide a sharp upper bound on the amount of coherence 
which can survive information transfer.
For perfect information transfer, 
all coherence vanishes.
This clearly proves that decoherence on a system is a mathematical
consequence of information transfer out of this system.
If, on the other hand, all states are left intact, then it follows 
that 
no information can be obtained.
This is the Heisenberg principle. 
Theorem \ref{offdiag} and 
corollary \ref{gencol} connect these two extremes 
in a continuous fashion; they form
a unified description of the Heisenberg
principle and the collapse of the wave function.

\end{itemize} 

\noindent \textbf{Acknowledgements} \\*
I would like to thank Hans Maassen for his invaluable guidance and advice.


\begin{thebibliography}{BGM}
\bibitem[AFL]{AFL} L. Accardi, A. Frigerio, Y.G. Lu, \textsl{`The Weak
Coupling Limit as a Quantum Functional Central Limit'}, 
Commun. Math. Phys.\textbf{131}, 537--570, (1990). 
\bibitem[AK]{AK} E. Arthurs, J. Kelly,
\textsl{`On Simultaneous Measurement on a Pair of Conjugate Observables'}, 
Bell. Syst. Tech. J. \textbf{44}, 725, (1965). 
\bibitem[BGM]{Bo}
L. Bouten, M. Gu\c{t}\u{a}, H. Maassen, \textsl{`Stochastic Schr\"odinger Equations'},
J. Phys. A. \textbf{37}, 3189--3209, (2004). 
\bibitem[Ha]{Ha} M. Hall, \textsl{`Prior Information: How to circumvent
the Standard Joint-Measurement Uncertainty Relation'}, Phys. Rev. A 
\textbf{69}, 052113, (2004).
\bibitem[He]{He} W. Heisenberg, \textsl{`\"Uber den anschaulichen Inhalt
der quantentheoretischen Kinematik und Mechanik'}, 
Z. Phys. \textbf{43}, 172--198, (1927). 
\bibitem[Hp]{Hp} K. Hepp, \textsl{`Quantum Theory of Measurement and 
Macroscopic Observables'}, Helv. Phys. Acta {\bf 45}, 237--248, (1972). 
\bibitem[Ho]{Ho} A. S. Holevo, \textsl{`Probabilistic and Statistical
Aspects of Quantum Theory'}, North Holland Publishing Company, (1982). 
\bibitem[Is]{Is} S. Ishikawa, 
\textsl{`Uncertainty Relations in Simultaneous Measurements for Arbitrary
Observables'},
Rep. Math. Phys. \textbf{29}, 257--273, (1991).
\bibitem[JB]{Luc} B. Janssens, L. Bouten, \textsl{`Optimal Pointers for 
Joint Measurement of $\sigma_x$ and $\sigma_z$ via Homodyne Detection'},
J. Phys. A. \textbf{39}, 2773--2790, (2006).
\bibitem[JM]{JM} B. Janssens, H. Maassen, \textsl{`Information Transfer
Implies State Collapse'},\\* {\tt www.arxiv.org/abs/quant-ph/0602140},
(2006).
\bibitem[JZ]{JZ} E. Joos, H. Zeh, \textsl{`The Emergence of Classical Properties
Through Interaction with the Environment'}, Z. Phys. B {\bf 59}, 223--243,
(1985).
\bibitem[Ke]{Ke} E. Kennard, \textsl{`Zur 
Quantenmechanik einfacher Bewegungstypen'}, Z. Phys. \textbf{44}, 
326--325, (1927).
\bibitem[Lu]{Lu} A. Lupa\c{s}cu  e.a. , \textsl{`High-Contast Dispersive
Readout of a Superconducting Flux Qubit using a Nonlinear Resonator'},
Phys. Rev. Lett. \textbf{96}, 127003, (2006).
\bibitem[Ne]{Ne} J. von Neumann, \textsl{`Mathematische Grundlagen der
Quantenmechanik'}, Springer-Verlag, (1932).
\bibitem[NC]{NC} M. Nielsen, I. Chuang, \textsl{`Quantum Computation and Quantum
Information'}, Cambridge University Press, (2000).
\bibitem[Oz]{Oz} M. Ozawa, \textsl{`Universally valid
Reformulation of the Heisenberg Uncertainty Principle on
Noise and Disturbance in Measurement'}, Phys. Rev. A \textbf{67}, 042105
(2003). 
\bibitem[Per]{Per} A. Peres, \textsl{`Quantum Theory: Concepts and Methods'}, 
Kluwer Academic Publishers (1993).
\bibitem[Ro]{Ro} H. Robertson, \textsl{`The Uncertainty Principle'}, Phys. Rev. 
\textbf{34}, 163--164, (1929).
\bibitem[Se]{Se}G. Sewell, \textsl{`On the Mathematical
Structure of Quantum Measurement Theory'}, Rep. Math. Phys. \textbf{56},
271--290, (2005).
\bibitem[Ta]{Ta} M. Takesaki, \textsl{`Theory of Operator Algebras I'},
Springer-Verlag, New York, (1979). 
\bibitem[We1]{We2} R. Werner, \textsl{`Optimal Cloning of Pure States'},
Phys. Rev. A, \textbf{58}, 1827--1832, (1998).
\bibitem[We2]{We} R. Werner, \textsl{`Quantum 
Information Theory -- an Invitation'}, Springer Tracts in Modern Physics
{\bf 173}, 14--57, (2001).
\bibitem[Zu]{Zu} W. Zurek, \textsl{`Environment-Induced Superselection
Rules'}, Phys. Rev. D {\bf 26}, 1862--1880, (1982).


\end{thebibliography}
\end{document}